\documentclass[aps,prl,twocolumn,showpacs,floatfix,10pt]{revtex4-1}
\usepackage{amsmath}
\usepackage{graphicx}
\usepackage{color}
\usepackage{epstopdf}
\usepackage{natbib}
\usepackage{xfrac}
\usepackage[normalem]{ulem}
\usepackage{multirow}

\begin{document}
\title{Importance of Schottky barriers for wide-bandgap thermoelectric devices}


\author{M.~Wais$^1$}
\email[]{michael.wais@tuwien.ac.at}
\author{K. Held$^1$}
\author{M. Battiato$^2$}
\affiliation{$^1$ Institute for Solid State Physics, TU Wien,  Vienna, Austria}
\affiliation{$^2$ Nanyang Technological University, 21 Nanyang Link, Singapore, Singapore}

\begin{abstract}
The development of thermoelectric devices faces not only the challenge of optimizing the Seebeck coefficient, the electrical and thermal conductivity of the active material, but also further bottlenecks when going from the thermoelectric material to an actual device, e.g., the dopant diffusion at the hot contact. We show that for large bandgap thermoelectrics another aspect can dramatically reduce the  efficiency of the device: the formation of Schottky barriers. 
Understanding the effect, it can then be fixed rather cheaply by  a two-metals contact solution. 
\end{abstract}

\date{\today}

\maketitle

\section{I. Introduction}

In recent years, vast scientific efforts have been invested in increasing the performance of thermoelectric devices \cite{Zlat}. Already a number of very attractive applications for sustainable energy solutions \cite{Snyder2008}  and technological applications such as Peltier coolers, heat pumps \cite{Dresselhaus2013}, microscopic generators \cite{PhysRevB.89.125103} and probes for quality control of solid state devices \cite{thermoelectric_imaging} are being pursued. But for  widespread applications more efficient thermoelectric devices are badly needed.

The development and optimization of a thermoelectric device is a task that requires tuning a number of parameters. The choice of a material which excels in interrelated and often competing thermoelectric properties  is imperative. That is, one is searching for materials with a large Seebeck coefficient which has, at the same time, a good electric and low thermal conductance. This has triggered an  intensive quest for new higher performing materials \cite{Dresselhaus07,Snyder2008,Yang16,BattiatoFeSb,Sun_CoSb3,GAYNER2016}.

However even highly promising materials can struggle to find application in commercial devices due to deficiencies in auxiliary properties that become important when going from the bulk thermoelectric material to an actual device.
Several such properties have been identified and discussed in the literature:  proneness to interdiffusion \cite{Uberuaga02,Hartmut05,Nam-Ho}, lack of chemical stability and resistance to oxidation \cite{Zhiyong12}, poor mechanical properties or low resistance to mechanical stresses \cite{Prokhorov11}, a too broad temperature dependence of the thermal expansion coefficient \cite{HUANG2006122}, a low 	resistance to heat or a low melting point, as well as toxicity. New thermoelectric materials such as oxide thermoelectrics \cite{Matsubara01,Shingo05,Ohta07,Lu10,Lin15} may excel over traditional semiconductors in many of the mentioned properties.


In this work we show that, for wide band-gap materials, an arguably even more important effect is the formation of Schottky barriers which can dramatically reduce the efficiency of the thermoelectric device. While the importance of Schottky barriers for semiconducting electronics is well established
 \cite{semicond}, its role  for thermoelectric devices
has been by-and-large overlooked in the literature. 
Neither is it 
taken into account in simulations for the thermoelectric efficiency \cite{Altenkirch,eng}, except sometimes indirectly  as a (constant) contact resistance of unknown origin \cite{Rowe96,Annapragada12,ASWAL201650,Kim17,note_schottInj}. 
Noteworthy and consistent with our findings, the experimental contact resistance
appears to be  particularly critical for wide-bandgap thermoelectric materials \cite{Yimin10,Yuhang16,Matsubara01}, making their performance consistently worse than expected \cite{ASWAL201650}. This is a very problematic issue, since it discourages the pursuit of otherwise very promising materials. 

In our paper, we compute these losses. Understanding that the formation of a Schottky barrier plays an essential role,  we  propose a simple and inexpensive way to mend the effect using a two-metals contact. Experimentally such a setup might have been achieved in some cases  by chance, when engineering the contacts of the thermoelectric device by trial-and-error.

Our calculations  use the standard charge and heat transport equations \cite{Battiatoprb17}, but
beyond  the dependence on the temperature considered before  we explicitly include also the
  dependence of all the transport properties on the chemical potential.
The latter is essential to describe the Schottky barrier
and has been considered for semiconductor electronics, but there in turn the temperature-dependence is not relevant and has been discarded. 

In the following, we first introduce the relevant equations. 
We then apply the treatment to two different materials for illustrative purposes.
We compare, for the small-bandgap thermoelectric material Bi$_2$Te$_3$, our treatment with the most commonly used estimations of the efficiency, showing how both standard estimations and more advanced ones \cite{eng,Kim17} overestimate the device efficiency.
Finally we apply our method to the large-bandgap thermoelectric SrTiO$_3$. We conclude that if the two branches of the thermoelectric device are formed by the same but differently doped active material, a single metal contact is insufficient to avoid the formation of a  Schottky barrier. 
We show that and how the contacts have to be composed by two different metals to prevent it.
The reader should notice however that several other sources of loss can critically compromise junctions, or the performance of thermoelectric devices.


\section{II. Theory and Method}

\begin{figure}[tb]
 \includegraphics[width=8.5cm]{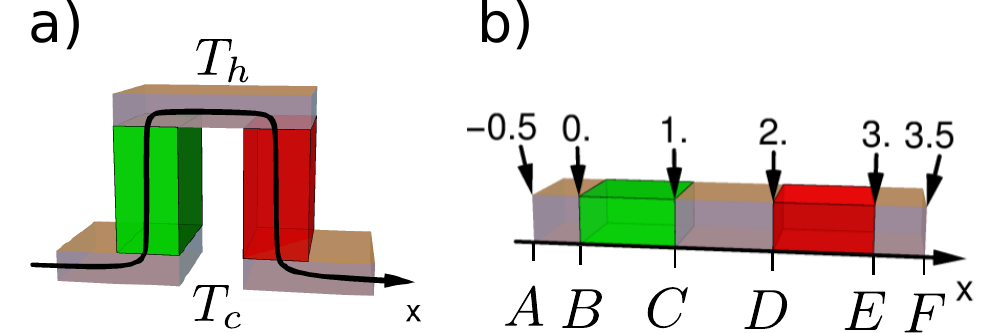}
 \caption{a) Schematic picture of a thermoelement. The red and green areas are thermoelectric materials and the grey areas are metals held at temperatures $T_h$ and $T_c$. b) Corners are neglected and the device is treated as an effective one dimensional device. The junctions are labeled from $A$ to $F$; positions are given in $mm$.
}\label{fig:thermelem}
\end{figure}

For the theoretical description, we assume an effective one-dimensional device consisting of two active regions (green and red areas in Fig.~\ref{fig:thermelem}) which are connected by metallic contacts held at different temperatures $T_h$ (hot side) and $T_c$ (cold side). The coordinate $x$ represents the position across the device. As already pointed out we maintain the full dependence of the transport coefficients on the temperature $T$ {\em and} the chemical potential $\mu$. The  Seebeck coefficient $S$, the charge density $\rho$,  the electrical $\sigma$, and thermal conductivity $\kappa$ depend on  the material considered as well as on  $T$  and $\mu$ (and doping):
\begin{align}
S (x)= S \left ( \mu(x),T(x) \right ) \textrm{, } 
&\rho (x)= \rho \left ( \mu(x),T(x) \right ) \textrm{,}  \label{eq:traspprop1} \\ \sigma (x) = \sigma \left ( \mu(x) ,T(x) \right) \textrm{, } & \kappa (x) = \kappa \left ( \mu(x) ,T(x) \right)\textrm{,} \label{eq:traspprop2}
\end{align}
where we have highlighted that these are  position dependent through the (yet to be determined) position dependence of $T(x)$ and  $\mu(x)$.



Within the active regions,  three equations are necessary to determine   $T(x)$,  $\mu(x)$, as well as the electrical potential  profile $\phi(x)$  \cite{Zlat}:
\begin{align}
&j = -\sigma(x) \left (  \phi '(x) - \frac{1}{e} \mu '(x)  +S(x) T'(x) \right ) \textrm{ ,}\label{eq:j1} \\
&\frac{j^2}{\sigma(x)} + \kappa'(x) T'(x) + \kappa(x)  T''(x) - T(x) j S'(x) = 0 \textrm{ ,} \label{eq:domeni}\\
&\phi''(x) = -  \frac{\rho(x)}{\epsilon_0 \epsilon} \; .\label{eq:poisson} 
\end{align}
Here, $j$ is the charge current, and $\epsilon_0 \epsilon$ the dielectric constant of the material and $e$ the absolute value of the electron charge. 
The first equation is the macroscopic charge transport equation where charge conservation is imposed by enforcing a spatially constant current. 
The second one is the so-called Domenicali equation \cite{domeni} which corresponds to energy conservation; and  Eq.~\eqref{eq:poisson} is the Poisson equation. 

Before addressing the problem numerically, we can further eliminate $\phi(x)$ in Eq.~\eqref{eq:j1} by means of Eq.~\eqref{eq:poisson} leading to the second-order differential equation
\begin{equation}
-\frac{1}{e} \mu''(x) = \frac{\rho(x)}{\epsilon_0 \epsilon} + \frac{j}{\sigma(x)^2} \sigma'(x) - \left(S(x) T'(x)\right)' \textrm{.} \label{eq:mu1}
\end{equation}
The differential Eqs.~(\ref{eq:domeni}) and (\ref{eq:mu1}) need to be solved self-consistently together with the material-dependent properties in Eqs.~\eqref{eq:traspprop1} and \eqref{eq:traspprop2}.
 This yields  $\mu(x)$ and $T(x)$  if we assign  the boundary conditions 
\begin{align}
& T(B) = T(E) = T_c \textrm{ , } T(C) = T(D) = T_h \\
&\mu(B) = \mu(C) = \mu(D) = \mu(F) = \mu_{\mathrm{F}} \; . \label{eq:boundarychempot}
\end{align}
Here, $\mu_\textrm{F}$ is the Fermi-level of the metal contacts \footnote{Note that $\mu_{\mathrm{F}}$ in reality deviate somewhat from the actual Fermi-level of the metal because of, e.g., impurities at the interface or charging of the first atomic layer on the metal side.}. 
The one above is a very good approximation in the usual case where the density of states of the metal is much bigger than that of the active material.

Once the equations have been solved, the efficiency of the thermoelectric device can be computed as
\begin{equation}
\eta =  \frac{P_{el}}{J_Q}=\frac{ \left (\bar \phi(F)-\bar \phi(A) \right ) j}{j_Q(D)-j_Q(C)} \textrm{ ,} \label{eq:eff}
\end{equation}
where $\bar \phi (x)\equiv  \phi (x) - \frac{1}{e} \mu (x)$ is the  electrochemical potential which is equivalent to the local voltage, and
\begin{equation}\label{eq:jQ1}
	j_Q (x)= S(x) T(x)  j - \kappa(x) T'(x) 
\end{equation}
 is the heat current.
To identify the maximum efficiency $\eta_{max}$ we solve the Eqs.\ for different operation conditions, characterized by different currents $j$.

For the examples below, 
the chemical potential- and temperature-dependent transport properties of Eqs.~\eqref{eq:traspprop1} and \eqref{eq:traspprop2} have been obtained using BoltzTraP \cite{boltztrap} with  bandstructures calculated from  density functional theory (DFT) using WIEN2K \cite{wien2k}. The doping is treated by assuming 
a rigid bandstructure and adding the dopant ionic charge to the charge expression in Eq.~\ref{eq:traspprop1}. 
To obtain the transport properties from BoltzTraP we need the relaxation time $\tau(T)$ which is determined by fitting the electrical conductivity to experimental values for a given carrier concentration. Furthermore the unit cell volume is needed in order to get the correct charge density. For the phononic part of the thermal conductivity we assume $\kappa_\textrm{ph} (T)= \alpha / T$ where the parameter $\alpha$ is again determined by fitting to experiments.

 We solve Eqs.~\eqref{eq:domeni} and \eqref{eq:mu1} numerically within the two active regions using finite elements with a non-uniform mesh, determining the position dependence of the transport properties self-consistently.

Below we compare our numerical results to the most common expressions employed in the literature to estimate the maximum efficiency of a thermoelectric element. For a device with constant transport coefficients the maximum efficiency can be calculated analytically and reads  
\begin{equation}
\eta(Z) = \eta_c \frac{\sqrt{1+Z T_m}-1}{\sqrt{1+Z T_m}+T_c / T_h}\label{eq:gen_eff1}
\end{equation}
where $ZT_m$  is the celebrated   figure of merit with $Z=\sigma S^2 / \kappa$; $T_c$ ($T_h$) the temperature at the cold (hot) side; $T_m$ the mean temperature i.e. $T_m = (T_h + T_c) /2$ and $\eta_c=(T_h-T_c)/T_h$ the Carnot efficiency.  

For real devices we will compare our numerical simulations with three common approximations: i) the efficiency given by Eq.~\eqref{eq:gen_eff1} with $Z$ evaluated at the mean temperature $T_m$, ii) the efficiency given by Eq.~\eqref{eq:gen_eff1} but with the temperature-averaged figure of merit $\eta_{ii} \equiv \eta \left (Z= \frac{1}{T_h-T_c} \int_{T_c}^{T_h}\mathrm d T \frac{S (T)^2 \sigma (T)}{\kappa (T)} \right ) $ and iii) the efficiency corresponding to the recently proposed engineering figure of merit $\eta_{eng}$ \cite{eng}. Note that our treatment is more general than (iii) and yields the latter if the temperature profile is linear (i.e., $T(x)=T_c+(T_h-T_c) x$ between $B$  and $C$ in Fig.~\ref{fig:thermelem} and analogously between $D$ and $E$) and if one assumes that $\mu(x)$ is everywhere at its equilibrium value. 

\section{III. Results and Discussion}
\subsection{A. Bi$_2$Te$_3$}

We first study the case of the widely used, narrow bandgap thermoelectric material, n- and p-doped bismuth telluride  (Bi$_2$Te$_3$; $n_{dop}= \pm 2.5 *10^{18} cm^{-3}$; $\tau(T)=\left ( 25.05*Exp[-T/93.65] + 1.11 \right ) *10^{-14}s$ \cite{abIniRel,Rel_Quin}; rhombohedral unit cell with $a=1.047nm$ and $c=3.048nm$ \cite{abIniRel}; $\kappa_{ph} = 540/T$ \cite{abIniRel}; $\epsilon=100$ \cite{bi2te3_diel}), operating between the temperatures $T_c=300K$, $T_h=600K$. We want to compare the effect of contacting the active regions with different metals, which we model by different contact potentials $\mu_{\mathrm{F}}$. 

\begin{figure}[tb]
 \includegraphics[width=8.5cm]{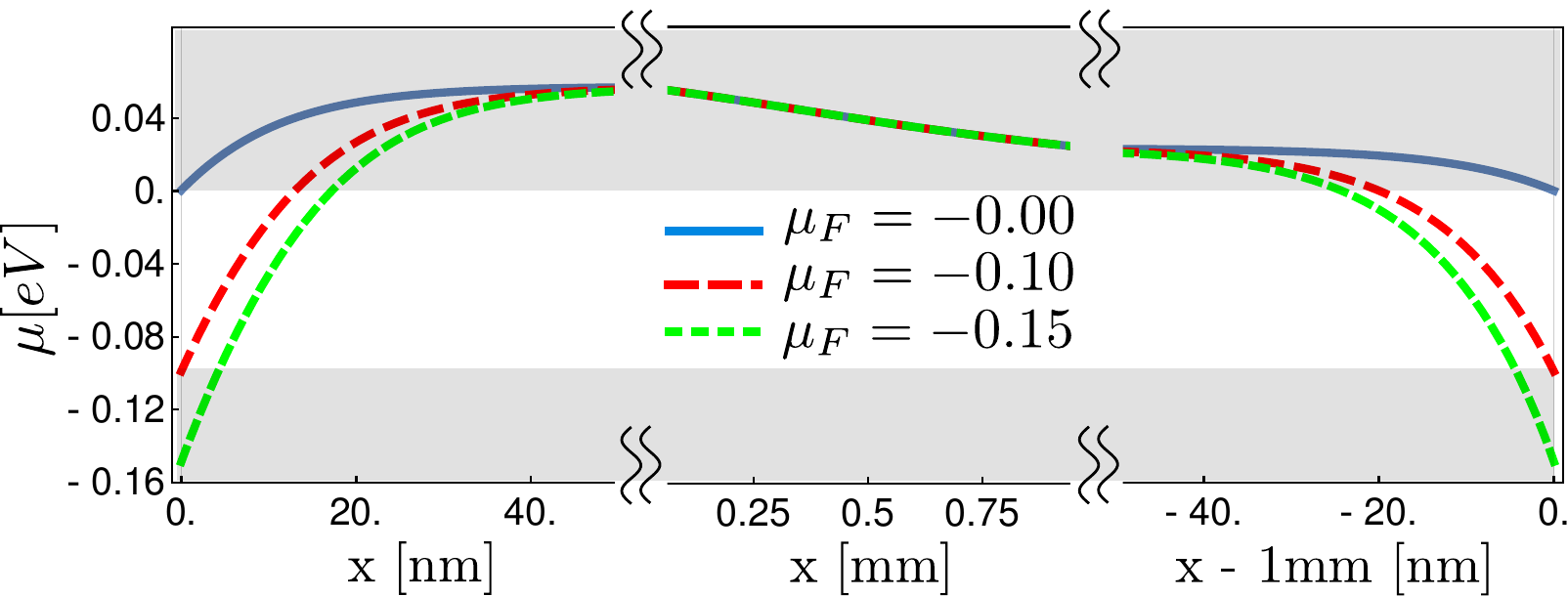}
 \caption{Chemical potential across the n-doped Bi$_2$Te$_3$ region for different chemical potential  $\mu_{\mathrm{F}}$ of the contact, $T_c=300K$ and $T_h=600K$. The current that gives the maximum efficiency (Tab.~\ref{tab:eff1}) is $j=3.2 \textrm{A/mm}^2$ for all cases. Note the zoom in (the different length scale) at the two interfaces. The bottom of the conduction band is taken as zero point. The gray areas mark the conduction and valence band.
}\label{fig:bi2te3_mu}
\end{figure}

In Fig.~\ref{fig:bi2te3_mu} we plot the position dependence of the chemical potential within the n-doped arm of the device for the three cases. We find that the chemical potential within the bulk hardly deviates from its equilibrium value. However, at the junctions, it is forced to the boundary value $\mu_{\mathrm{F}}$ (left and right section of Fig.~\ref{fig:bi2te3_mu}). 
In spite of the explicit chemical potential dependence of the transport properties, the change in the transport coefficients at the boundaries is not important enough to appreciably affect the efficiency. This is because the bandgap ($E_{gap} \approx 0.11$eV\cite{bite_tau}) is not much larger than $T_c$ ($0.026$eV) so that there remain temperature-induced charge carriers even if the chemical potential lies within the gap, and the deviation of the chemical potential is anyhow restricted to a small region of ${\cal{O}}(10)$ nm.

\begin{center}
\begin{table}[]
\centering
\begin{tabular}{c|ccc|c|c|c|}
\cline{2-7}
               & \multicolumn{1}{c|}{$\eta_i$} & \multicolumn{1}{c|}{$\eta_{ii}$} & \begin{tabular}{c}  $\eta_\texttt{eng}$ \\  \cite{eng}  \end{tabular}      & \multicolumn{3}{c|}{$\eta_\textrm{sim}$}   \\ \hline
\multicolumn{1}{|c|}{\multirow{2}{*}{Bi$_2$Te$_3$}} &                               &                                  &        & $ \mu_{\mathrm{F}} = 0.$ & $ \mu_{\mathrm{F}} = -0.1$ & $ \mu_{\mathrm{F}} = -0.15$ \\ \cline{2-7} 
\multicolumn{1}{|c|}{}                              & \multicolumn{1}{c|}{5.34\%}   & \multicolumn{1}{c|}{5.11\%}      & 4.94\% & 4.85\%                               & 4.85\%                                 & 4.85\%                                  \\ \hline
\multicolumn{1}{|c|}{\multirow{2}{*}{SrTiO$_3$}} &                               &                                  &        & $ \mu_{\mathrm{F}} = 0.$ & $ \mu_{\mathrm{F}} = -0.5$ & $ \mu_{\mathrm{F}} = -1.$ \\ \cline{2-7} 
\multicolumn{1}{|c|}{}                              & \multicolumn{1}{c|}{1.64\%}   & \multicolumn{1}{c|}{ 1.58\%}      & 1.52\% & 1.57\%                               & 1.1\%                                 & $1.61 *\!10^{-5}$\%                            \\ \hline
\end{tabular}
\caption{Comparison of the different efficiency estimations and the simulations.}
\label{tab:eff1}
\end{table}
\end{center}

In Tab.~\ref{tab:eff1} we compare the different estimations of the efficiency for the n-doped leg with the one obtained in this work. The estimations $\eta_i$ and $\eta_{ii}$ are known to overestimate the efficiency, which is confirmed here. The efficiency estimation from Ref. \cite{eng} agrees better with the simulated values. The remaining deviation is due to the non-linear temperature distribution in the real system.

\subsection{B. SrTiO$_3$}
The situation is entirely different in doped strontium titanate (SrTiO$_3$; $n_{dop} =\pm 10^{20}cm^{-3}$; $\tau(T)= \left (10.125 * Exp[-T/493.261] - 1. \right )*10^{-14}s$ obtained by fitting to data in Ref. \cite{srtio_relTime}; $\kappa_{ph}=1600/T $ obtained by fitting to data in Ref. \cite{srtio_allg}; cubic unit cell with $a=0.3905nm$ \cite{srtio_allg}; $\epsilon \approx 288$ at room temperature\cite{srtio_diel}), which is a high bandgap material with admirable thermoelectric properties \cite{srtio_allg}. We again compare the running conditions for three different values of $\mu_{\mathrm{F}}$. In all three cases the temperature slope is almost linear (not shown). When the metal's Fermi level lies at the bottom of the conduction band, we find excellent agreement between $\eta_i$, $\eta_{ii}$, $\eta_{eng}$ and the simulation (first four columns in Tab.~\ref{tab:eff1}). This is understandable since the transport properties do not vary strongly with temperature so all the approaches become equivalent.

\begin{figure}[tb]
 \includegraphics[width=8.5cm]{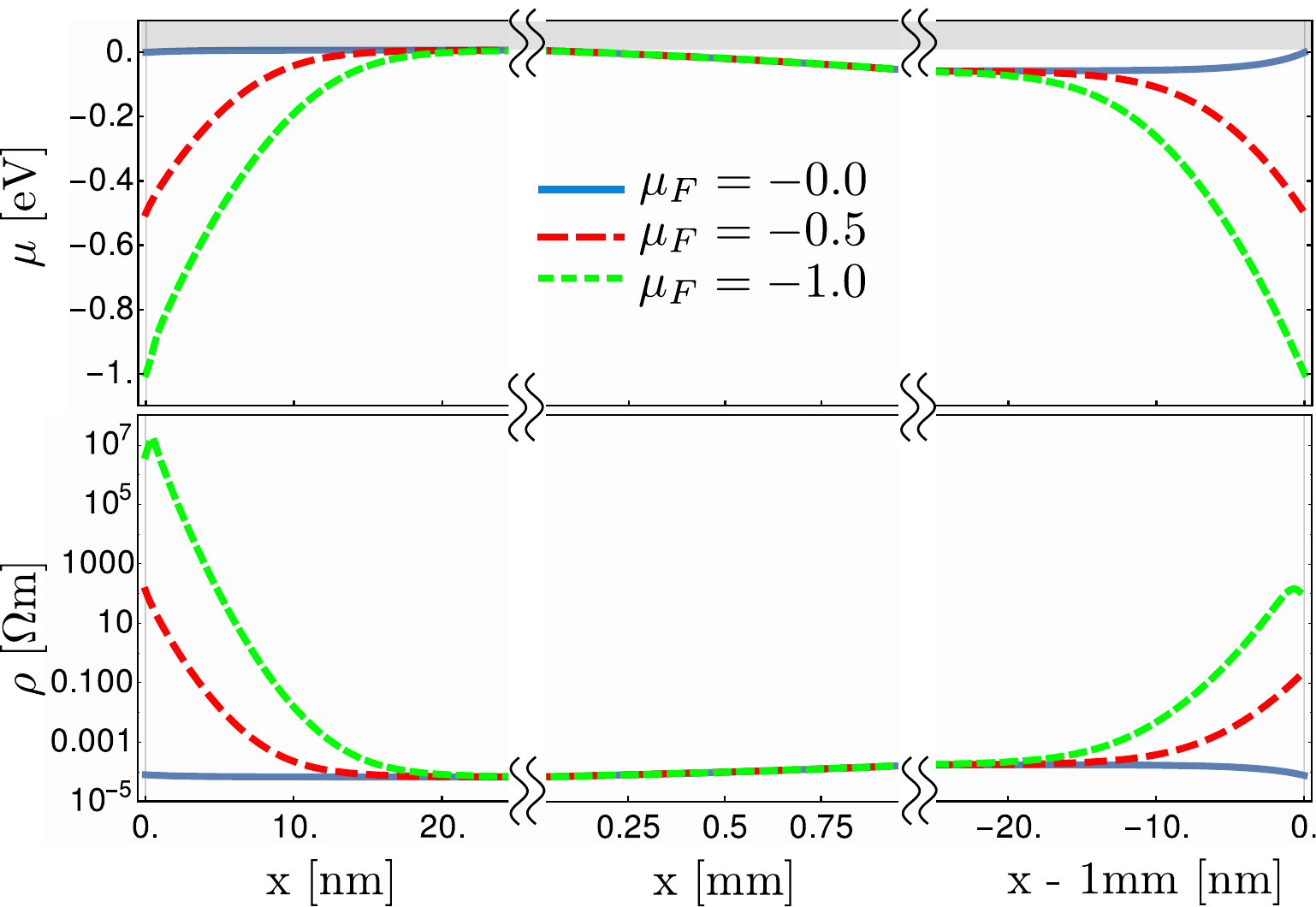}
 \caption{Chemical potential as in Fig. \ref{fig:bi2te3_mu} and the corresponding specific resistivity of n-doped SrTiO$_3$. The applied temperatures are $T_c=400K$ and $T_h=700K$. The currents that give the maximum efficiency (Tab.~\ref{tab:eff1}) are: $j=400  \textrm{mA/mm}^2$ (blue), $j=280  \textrm{mA/mm}^2$ (red) and $j=5.6 \mu \textrm{A/mm}^2$ (green). Note that the band-gap of the simulation is smaller than in reality ($E_{gap} \approx 1.9eV$).
}\label{fig:srtio3_mu}
\end{figure}

On the other hand the efficiency dramatically drops if the metal's Fermi level lies deep within the bandgap. The reason for this is the formation of a Schottky barrier and an associated depletion region (widened by the large dielectric constant in SrTiO$_3$). This creates a highly resistive region close to the interface where $\mu(x)$ lies within the now much larger bandgap (see Fig.~\ref{fig:srtio3_mu}). This depletion region makes the total electrical resistivity of the layer comparable to the resistivity of the rest of the device \footnote{Note that Eqs. \eqref{eq:j1} and \eqref{eq:domeni} describe diffusive transport. Corrections coming from ballistic transport are neglected.}. As Table \ref{tab:eff1} demonstrates, the Schottky barrier can even completely suppress the thermoelectric efficiency of a SrTiO$_3$-based device. Because of its much smaller band-gap, this effect is negligibly small for Bi$_2$Te$_3$.

\begin{figure}[tb]
 \includegraphics[width=0.45\textwidth]{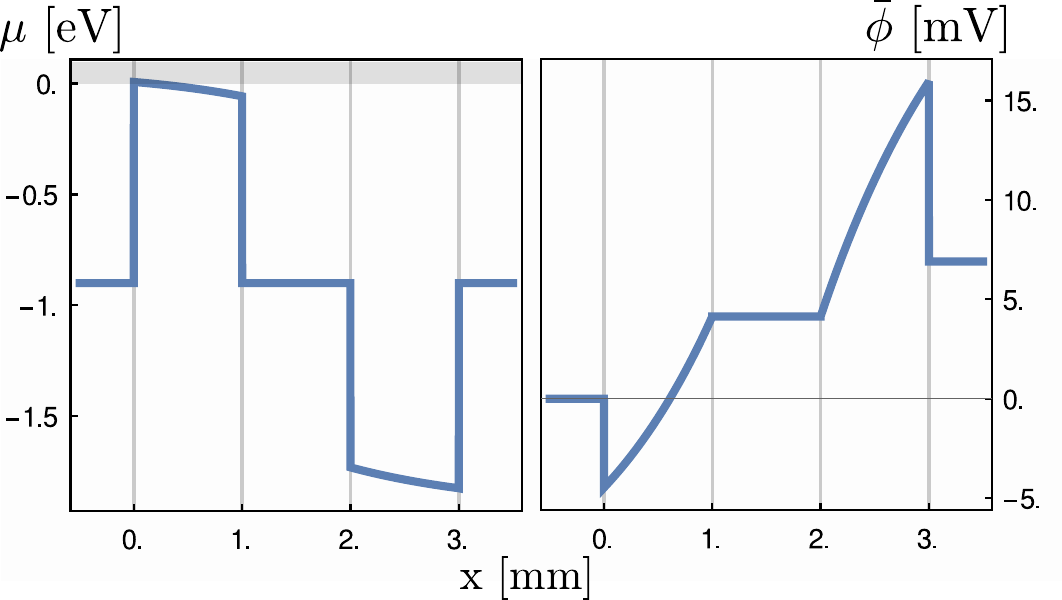}
 \caption{Chemical potential and electrochemical potential along the  whole  SrTiO$_3$-based device for metallic contacts corresponding to $\mu_{\mathrm{F}}=-0.9\,$eV and $j=25 \mu \textrm{A/mm}^2$ which gives an efficiency of $\eta = 6.7 *10^{-5}\%$. At the cold contacts ($x=0$ and 3$\,$mm) we observe a drop in the electrochemical potential due to the increased resistivity.
}\label{fig:srtio3_mu_elch}
\end{figure}

The right panel of Fig.~\ref{fig:srtio3_mu_elch} shows the electrochemical potential throughout the whole device. 
From left to right, $\bar \phi (x)$ is almost constant within the (highly conducting) metal contact. At the interface  ($x=0$) there is very sharp drop (note that the depletion region is much smaller than the dimension of the active element). Within the thermoelectric element a voltage is built. At the hot contact ($x=1\,$mm)  there is no discernible drop. This is due to the natural dependence of the conductivity of a semiconductor on the temperature. A higher temperature allows more thermally excited carriers, even when the chemical potential is within the bandgap, dramatically reducing the resistance of the Schottky barrier. A similar behavior is observed at the other branch of the device at $x=3\,$mm. The voltage drops at the cold interfaces lead to a much smaller voltage difference across the  thermoelectric device (from  $x=-0.5$ to $3.5\,$mm), and hence a loss  of produced power.   

Notice that, due to the non-equilibrium position of the chemical potential at the junctions, the Seebeck coefficient and the thermal conductivity change as well. However, this has negligible influence on the efficiency since the change of $\mu(x)$ is restricted to a tiny region and the Seebeck coefficient is affected almost linearly and the phononic part of the thermal conductivity is not influenced at all. The reader should notice how the usual estimations for the efficiency dramatically fail in recognizing this behavior (Tab.~\ref{tab:eff1}).

\subsection{C. Two-metal contact solution}
\begin{figure}[t]
 \includegraphics[width=0.45\textwidth]{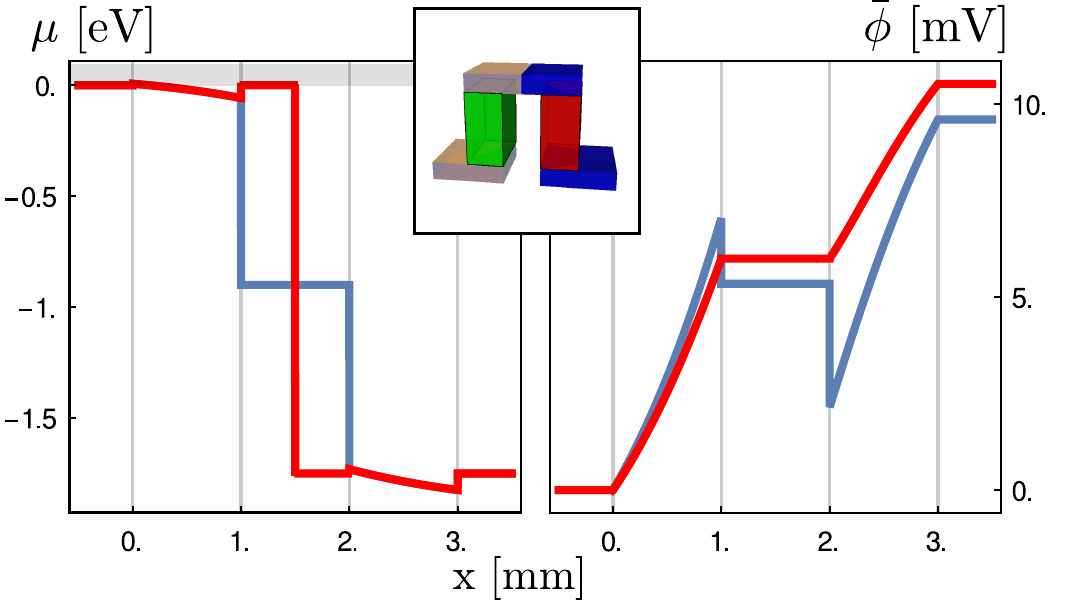}
 \caption{Same as Fig. \ref{fig:srtio3_mu_elch} but with different metallic contacts for each leg. Two cases are shown, one where only the cold contact has been optimized (blue) and one where both contacts are optimized (red). One can see that there is a drop in the electrochemical potential at the hot side for the blue case and no drop at all for the red one. The currents and corresponding efficiencies are $j=150 \textrm{mA/mm}^2$, $\eta = 0.7 \%$ (blue) and $j=250 \textrm{mA/mm}^2$, $\eta = 1.22 \%$ (red).  The inset shows a schematic picture of the device where both contacts are optimized (red case). 
}\label{fig:srtio3_mu_elch2}
\end{figure}

The dramatic drop in efficiency due to the formation of a Schottky barrier is therefore important in large-bandgap materials and especially at the cold side. The first obvious approach to the problem would be to select a metal with an appropriate chemical potential. However often the two legs of a thermoelectric device consist of the same material but with opposite doping. In that case the metal used for the cold side would contact both the n- and the p-doped active regions. If the Fermi level is chosen to be optimal for one side, it will instead be extremely disadvantageous for the other side.  This can be bypassed by using two different metals as shown in Fig.~\ref{fig:srtio3_mu_elch2} (notice that no depletion regions are created at a metal-metal junction).  Another  possible solution would be to highly dope the regions around the cold junctions, leading to a faster decay of the depletion regions \cite{semicond}. The reader might be led to believe that only the  optimisation of the cold contact is required. However, our simulations show that that is not the case. By the cold contact optimisation the device can now run at higher currents, where usually higher efficiencies can be achieved. However, the presence of a Schottky barrier at the hot contact (even if less effective than the one at the cold side) still imposes important voltage losses and prevents reaching the maximum efficiency (Fig. \ref{fig:srtio3_mu_elch2}).

\section{IV. Conclusion}
In conclusion we have performed simulations of thermoelectric generators with the full dependence of the transport coefficients on temperature and chemical potential. We found that one cannot always neglect the explicit dependence of the transport properties on the chemical potential in large bandgap materials because a disadvantageous position can lead to the formation of Schottky barriers which completely destroys the performance of the thermoelectric generator. We propose  to use two different metals, separately optimised for the two ($n$- and $p$-doped) branches. The advantage of properly engineering the contacts is critical for large-bandgap thermoelectric materials, while requiring quite inexpensive adjustments of the design of the device. 

Finally,  we would like to remind the reader of Mahan's famous $10 k_BT$ formula, i.e., that the thermoelectric  material should have a gap size  of 10 times the operating temperature. However, the actual calculation \cite{Mahan89,Sofo94}  only gives a lower bound, i.e., the gap should be larger than  $10 k_BT$ but  no upper bound ($ZT$ still increases insignificantly). One might speculate that at least one important reason why larger bandgap materials typically underperform, making the lower bound also the upper bound, is the formation of Schottky barriers in wide-bandgap thermoelectrics.

\textit{Note added:} After submission we learned of a somewhat related work \cite{snyder_unpub} that models a depletion region at a grain boundary by a constant shift of the chemical potential in a region of fixed width. 

\section{Acknowledgement}
The authors wish to thank V.~Zlatic and E. Bauer for discussions and  acknowledge financial support by the European Research Council/ERC through Grant Agreement No.~306447 and by the Austrian Science Fund (FWF) through SFB ViCoM F41 and graduate school W1243 Solid4Fun. M. Battiato acknowledges the Austrian Science Fund (FWF) through Lise Meitner position M1925-N28 and the Nanyang Technological University, NAP-SUG for funding.


\begin{thebibliography}{51}
\bibitem [{not()}]{note_schottInj}%
  \BibitemOpen
  \href@noop {} {}\bibinfo {note} {To the best of our knowledge the only actual
  calculations of thermoelectric devices including Schottky barriers beyond a
  mere contact resistance of arbitrary origin, are in the context of high
  carrier infection \cite{enFiltGrain,schottky_inj1, schottky_inj2}. The aim is
  to boost the electron-hole asymmetry to increase the thermopower (but the
  authors concluded that that strategy does not work). We highlight that that
  treatment works only in the ballistic regime, and therefore is not applicable
  to devices larger than $\mu m$. Finally that treatment disregards the
  temperature profile within the thermoelectric, and, if applied to a whole
  device, it would fail to provide the physical insight that our work
  reaches.}\BibitemShut {Stop}%
\end{thebibliography}
\end{document}